\documentclass[journal, table]{IEEEtran}
\usepackage{amsmath, amssymb, booktabs, bm, cite, color, lipsum, siunitx, mathdots, multirow, textcomp, verbatim, xcolor}
 \usepackage{microtype}
\usepackage{threeparttable}

\usepackage{longtable}
\usepackage{graphicx}
\graphicspath{{figures/}}

\ifCLASSOPTIONcompsoc
  \usepackage[caption=false,font=normalsize,labelfont=sf,textfont=sf]{subfig}
\else
  \usepackage[caption=false,font=footnotesize]{subfig}
\fi

\usepackage{mdwmath}
\usepackage{eqparbox}
\usepackage{url}
\usepackage{algorithm}
\usepackage{algorithmicx}
\usepackage{algpseudocode}

\usepackage{booktabs} 
\usepackage{multirow} 
\usepackage{lipsum}   
\usepackage{tikz}
\usepackage{siunitx}
\usepackage{tabularx}
\usepackage{makecell}
\usepackage{booktabs}
\usepackage{multirow}

\usepackage{etoolbox}
\usepackage{hyperref}
\newtoggle{hl}
\togglefalse{hl}
\iftoggle{hl}{%

}{%

}

\begin{document}
\bstctlcite{IEEEexample:BSTcontrol}
\title{Multi-site Radar Systems for High-Precision \\ Indoor Positioning and Tracking}
  \author{Lang~Qin,~\IEEEmembership{Student Member,~IEEE,}
      Mandong~Zhang,
      Wenting~Song,
      Xiaohu~Wu,~\IEEEmembership{Senior Member,~IEEE,} \\
      Zhiqiang~Huang,~\IEEEmembership{Member,~IEEE,}
      and~Xiaoguang~Liu,~\IEEEmembership{Senior Member,~IEEE}

\thanks{Manuscript received January 21, 2026. This work was supported by the Shenzhen Science and Technology Program under Grant (JCYJ20230807091814030, JCYJ20220818100408018, and 20231115204236001), the National Natural Science Foundation of China under Grant (62471211 and 32371992), and in part by the Guangdong Basic and Applied Basic Research Foundation under Grant (2025A1515011109 and 2024A1515011902). Corresponding authors: Zhiqiang Huang and Xiaoguang Liu.}
\thanks{L.~Qin is with the Microelectronics Thrust, Hong Kong University of Science and Technology (Guangzhou) (HKUST(GZ)), as a Ph.D. student, and with the School of Microelectronics (SME), Southern University of Science and Technology (SUSTech), Shenzhen, China, as a visiting scholar.}
\thanks{M.~Zhang, W.~Song, and X.~Liu are with SME, SUSTech, Shenzhen, China.}%
\thanks{Z.~Huang is with the Microelectronics Thrust, HKUST(GZ), Guangzhou, China.}
\thanks{X.~Wu is with Peng Cheng Laboratory, Shenzhen, China.}}


\maketitle

\begin{abstract}
This paper introduces a high-precision indoor positioning and tracking method that utilizes multi-site single-input single-output (SISO) radar systems. We propose a novel velocity synthesis-assisted (VSA) localization algorithm that iteratively refines target position estimates within range bins by fusing radial velocity measurements from multiple radars. This approach ensures enhanced accuracy in both velocity and position estimation. Moreover, the inherent geometric constraints introduced by velocity synthesis enable the proposed algorithm to remain robust under low signal-to-noise ratio (SNR), severe multipath propagation, and large synchronization latency. Notably, our method eliminates the use of multiple-input-multiple-output (MIMO) configurations and stringent phase synchronization requirements, substantially reducing hardware complexity while maintaining high positioning accuracy. We define standardized reference trajectories to facilitate a comprehensive and reproducible performance evaluation. Extensive simulations and experimental validations demonstrate that our multi-site radar systems achieve centimeter-level tracking accuracy for human subjects, outperforming existing methods in complex trajectory tracking.
\end{abstract}

\begin{IEEEkeywords}
Indoor positioning, multi-site radar, velocity synthesis, human tracking, extended Kalman filter (EKF). 
\end{IEEEkeywords}

%
\IEEEpeerreviewmaketitle


\section{Introduction}

\IEEEPARstart{W}{ith} the rapid development of smart homes, industrial automation, and medical monitoring, the demand for high-precision indoor positioning and tracking technology has grown significantly. Traditional positioning methods, such as global positioning system (GPS) and Beidou, fail in indoor environments due to severe signal attenuation and multipath effects. Indoor localization can be broadly categorized into active (tag-based) and passive (tag-free) localization. Various tag-based indoor positioning methods exist, including radio-frequency identification (RFID)~\cite{Scherhufl_2013_RFID,Paolini_2019_RFID}, secondary radar~\cite{Roehr_2008_SecondaryRadar,Dobrev_2019_Locating}, ultrawideband (UWB)~\cite{Zhang_2010_Real-Time,Mahfouz_2008_Investigation}, etc. The premise of using these methods is that users must physically carry positioning tags, which can be inconvenient in many scenarios. 

Tag-free positioning approaches, on the other hand, do not require a physical tag attached and rely on non-contact methods for positioning.  For example, vision-based techniques, such as stereo-vision, time-of-flight (ToF), and structured light systems~\cite{Islam2020Stereo,morar2020comprehensive,silberman2011indoor}, are widely used to provide accurate location estimation. In many cases, vision-based positioning systems have high enough imaging resolution to provide further capabilities such as human identification and gait recognition~\cite{singh2018vision,wang2010review}. However, due to privacy concerns, vision-based solutions are not popular in home and many business settings~\cite{2021_IoT_Alam,li_Indoor_2023}. At the same time, radio frequency (RF)-based techniques have seen extensive research efforts in recent years, such as commercial wireless fidelity (Wi-Fi)~\cite{wang2015understanding,Yen_2022_3-D}, impulse radio ultra-wideband (IR-UWB) radar~\cite{guo_2024_uwtracking,Zheng_Catch_2023} and millimeter-wave (mmW) radar~\cite{canil_ORACLE_2024,li_Sequential_2021}.  
 
Radar sensors are an important sensing modality for indoor positioning~\cite{sesyuk2024radar}. Compared to cameras, microwave and mmW radars rely on electromagnetic scattering to retrieve spatial and kinematic information rather than capturing optical imagery. This fundamental difference in sensing modality makes radar a less intrusive solution for indoor applications. In addition, radars are capable of measuring speed/velocity which is critical information in applications such as gait/gesture recognition~\cite{li_Sequential_2021,Sun2021Gesture} and fall detection~\cite{Ding2023Fall,mercuri2023biomedical}. 
Compared to commercial Wi-Fi devices, microwave and mmW radars are not subject to the same hardware limitations that lead to significant carrier frequency offsets due to environmental changes, and therefore offer higher positioning accuracy.
Compared to IR-UWB radar, which is often favored for through-wall and non-line-of-sight (NLOS) scenarios due to its extremely high time resolution, mmWave radar offers superior spatial resolution and high-precision Doppler velocity measurements.


\begin{figure*}
    \centering
    \includegraphics{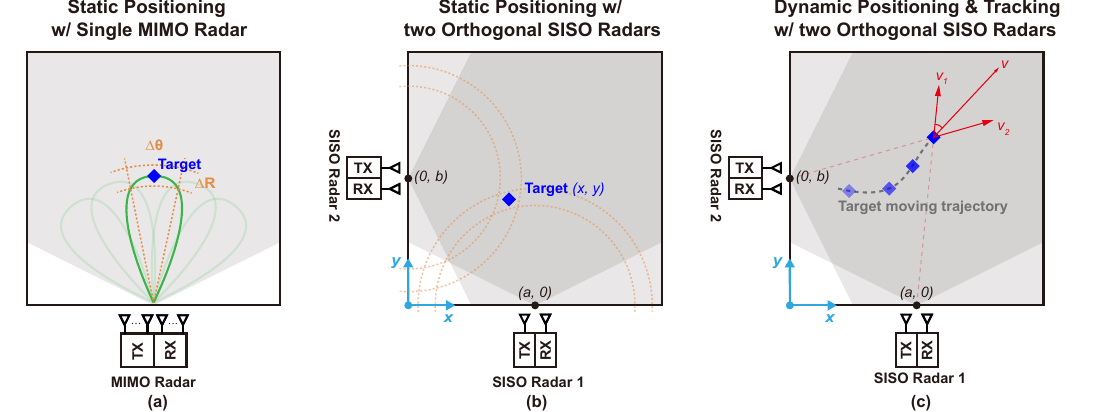}
    \caption{\ Indoor high-precision target single positioning. (a) Single MIMO radar positioning~\cite{Feng_TMTT2021}. (b) Multi-site SISO radar systems for static target positioning. (c) Multi-site SISO radar systems for dynamic target positioning and tracking.
    }
    \label{fig:main}
\end{figure*}

Spatial resolution is an important metric in radar systems. The radial or range resolution $\Delta R$ of a radar is primarily determined by occupied radio-frequency bandwidth $B$ of the radar signal and is given approximately by $\Delta R = c/2B$, where $c$ is the speed of light. Off-the-shelf radar sensors working at mmW frequencies can provide centimeter-level range resolution~\cite{SocionextSC1220AT2,TIR6843}. The angular resolution of a microwave/mmW radar is provided by either a phased array configuration or a multiple-input-multiple-output (MIMO) configuration as shown in Fig.~\ref{fig:main}(a). For an effective two-dimensional aperture of $N\times N$ antennas, the angular resolution in radians is approximately $\Delta \theta = 0.886\lambda/D$, where $D$ is the directivity of the effective array. Several MIMO radar systems for indoor applications have been demonstrated~\cite{Feng_TMTT2021,Su_SIL_2021,liRobustAccurateFMCW2024a,Song_TMTT2025}. For example, a $6\times6$ MIMO radar system is demonstrated in \cite{Feng_TMTT2021} with an angular resolution of $17^{\circ}$ at $2.45$ \,GHz which translates to a cross range resolution of $0.6$\,m at a distance of $1.8$\,m. Achieving very high angular resolution so that cross-range resolution can approach that of the range resolution requires very large and potentially impractical array sizes.

To address the inherent resolution limitations of single radar systems, multi-site radar systems have been proposed as a solution. Currently, strategies for implementing radar systems synchronization are generally divided into two categories: hardware connection and software connection. The first category relies on cables to realize bistatic measurements and synchronization. Radar networks using frequency-division multiplexing (FDM)~\cite{frischen_Cooperative_2017} or distributed monostatic/bistatic MIMO constellations~\cite{nguyen_High_2021,Fenske_Constellation_2024} have been explored. While these custom systems demonstrate excellent position and tracking precision, their practical deployment is severely constrained by the need for complex cabling, making them less suitable for consumer-level indoor applications. Another method uses multiple independent monostatic radar nodes to form multi-site radar systems by networks connection. The removal of physical cabling results in increased synchronization latency. Therefore, each radar node typically rely solely on monostatic measurements.

In addition to synchronization strategy, the effective fusion of data from spatially distributed radars presents another challenge. Existing methods can be categorized into two approaches: learning-based~\cite{2023_TGRS_learning,2023_TRS_learning,9454972} and geometry-based~\cite{canil_ORACLE_2024,wuUserIdentificationCollaborative2025,li_Sequential_2021,Zheng_Catch_2023,guo_2024_uwtracking}. Typical learning-based approaches map raw radar signatures directly to indoor positions by using end-to-end deep learning. Although promising, these methods often require extensive training datasets and lack model interpretability. Furthermore, significant errors may arise when the testing environment changes drastically. In contrast, geometry-based approaches offer superior generalization capabilities. 
For low-cost SISO radars, a conventional approach uses trilateration for positioning as shown in Fig.~\ref{fig:main}(b), followed by an extended Kalman filter (EKF) for tracking as shown in Fig.~\ref{fig:main}(c)~\cite{guo_2024_uwtracking,Zheng_Catch_2023}. However, distance-only measurements provide limited geometric constraints, which may result in multiple solutions or high geometric dilution of precision (GDOP), especially in software connection systems~\cite{10539071,Mugil2024EfficientMF}.

To increase geometric constraints, Zeng et al.~\cite{Zeng_2021_Massive} demonstrated that incorporating synthesized velocity measurements from massive MIMO into a state-space estimator can significantly improve localization and tracking accuracy. However, such velocity synthesis relies heavily on massive MIMO infrastructure, rendering it unavailable for typical low-cost consumer-grade indoor applications. Guo et al.~\cite{guo_2024_uwtracking} used historical trajectory information to constrain the tracking search angle. However, such methods rely heavily on the continuity of motion and often fail to adapt to abrupt maneuvering, such as sharp turns or rapid acceleration, leading to tracking divergence. Thus, a key challenge lies in designing a method using cable-free, low-cost indoor tracking systems suitable for consumer applications, which avoids the use of large MIMO arrays and custom hardware.

In summary, the main contributions of this work are as follows:
\begin{enumerate}

    \item High-precision indoor human positioning and tracking are achieved using two cable-free, low-cost SISO mmWave FMCW radars. Unlike conventional systems that rely on complex MIMO channels or strict hardware-level synchronization, our multi-site radar systems require only two orthogonal flexibly deployed, low-cost SISO radars.

    \item Building upon this multi-site radar systems, we propose a velocity synthesis-assisted (VSA) algorithm. This method incorporates radial velocity measurements as geometric constraints within a grid-based search space, thereby enabling precise target localization and velocity vector synthesis. We conducted extensive Monte Carlo simulations to evaluate the algorithm's robustness under non-ideal conditions, including low signal-to-noise ratio (SNR), severe multipath interference, and inter-radar synchronization latency. The results demonstrate that the proposed VSA algorithm realizes target positioning and tracking while maintaining centimeter-level accuracy.
    

    \item The algorithm was validated through experiments in both single-subject and dual-subject indoor scenarios. We implemented the systems using commercial off-the-shelf (COTS) millimeter-wave radars to avoid the systematic errors introduced by custom hardware systems. Experimental results indicate that in complex trajectory scenarios, our approach outperforms both single MIMO systems and recent state-of-the-art (SOTA) fusion algorithms. This confirms the feasibility of deploying high-precision, privacy-preserving indoor tracking solutions with reduced hardware cost and complexity.

\end{enumerate}

\section{Review of Multi-site Radar Positioning and Tracking Methods}

\subsection {Multi-site Radar Synchronization}

\begin{figure}
    \centering
    \includegraphics{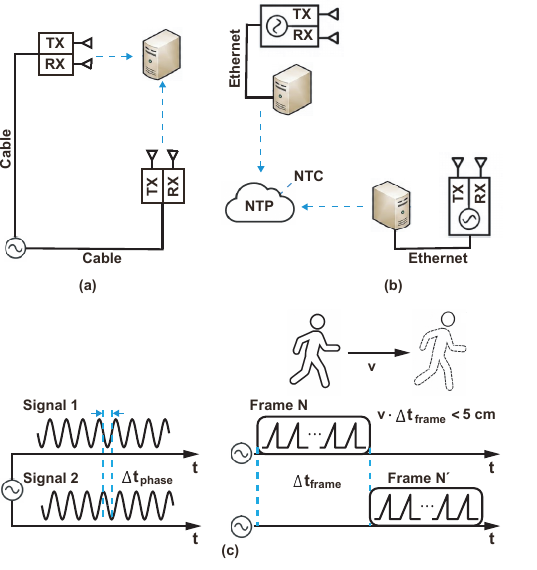}
    \caption{\ Multi-site radar synchronization. (a) Coherent solution.~\cite{nguyen_High_2021}. (b) Noncoherent NTP solution. (c) The comparison of synchronization precision requirements: coherent vs. non-coherent radar.}
    
    \label{fig:multi-site}
\end{figure}

Time synchronization is a foundational requirement in multi-site radar systems to ensure accurate data fusion. As mentioned in introduction, depending on different application requirements, time synchronization methodologies are generally classified into two categories: hardware connection and software connection. The hardware connection can serve two functions: phase synchronization and trigger synchronization. The former has a shared clock source across all radar nodes to ensure strict temporal and phase alignment, thereby referred to as coherent synchronization. Fig.~\ref{fig:multi-site}(a) depicts a coherent solution, where radars share a common physical clock source via cabling. In contrast, trigger synchronization entails only a synchronized start time, which is typically characterized as non-coherent synchronization. Software solutions can only realize trigger, which often use networks. Precision Time Protocol (PTP)~\cite{IEEE1588-2019} and Network Time Protocol (NTP)~\cite{RFC5905} are representative examples. The former can achieve nanosecond-level trigger synchronization but requires expensive dedicated networking equipment, whereas the latter provides only millisecond-level trigger synchronization without the need for additional network hardware, making it more suitable for consumer-grade products. Sharing a similar deployment philosophy with~\cite{canil_ORACLE_2024}, our system adopts a NTP-based synchronization architecture to ensure the distributed network remains lightweight and viable for practical indoor tracking, which is shown in Fig.~\ref{fig:multi-site}(b).

Coherent systems align the carrier wave phases of the signals, minimizing the phase offset $\Delta t_{phase}$ shown in the left panel of Fig.~\ref{fig:multi-site}(c). While this approach allows lower synchronization error, it imposes strict hardware constraints and limits the physical flexibility of the deployment. In contrast, for non-coherent synchronization, the focus shifts from carrier phase alignment to the alignment of data frames or pulses, represented by the offset $\Delta t_{frame}$ in the right panel of Fig.~\ref{fig:multi-site}(c). Typically, the synchronization error for coherent systems is on the order of picoseconds (ps), whereas for standard NTP-based systems, it is generally on the order of milliseconds (ms). For applications such as indoor human target tracking, the necessity of phase-level synchronization must be weighed against the motion characteristics of the target. Assuming a typical human walking speed of approximately $v \approx \SI{1.0}{m/s}$, even in very poor network environments, a temporal misalignment of \SI{50}{\text{ms}} results in a spatial displacement of $\Delta d = \SI{1.0}{m/s} \times \SI{50}{\text{ms}} = \SI{5.0}{\text{cm}}$. This displacement of \SI{5.0}{\text{cm}} is comparable to the theoretical range resolution of typical commercial millimeter-wave radars, while remaining significantly smaller than the characteristic dimensions of the human body. Consequently, a millisecond-level synchronization error is deemed acceptable for human-scale tracking scenarios.

Based on this analysis, this system adopts the practical non-coherent alignment scheme shown in Fig.~\ref{fig:multi-site}(b). The clocks of the multi-site radar nodes are triggered via Ethernet to a national timing center (NTC) server using NTP. Upon system initialization, the radars generate and exchange initial timestamps to establish a common temporal baseline with millisecond precision, ensuring that the frames captured by different nodes can be effectively associated in the temporal domain. It is worth noting that non-coherent architectures have been widely used in indoor positioning and tracking applications in IR-UWB radar systems due to their implementation simplicity~\cite{li_Sequential_2021,Zheng_Catch_2023,guo_2024_uwtracking}. While IR-UWB radar systems are sensitive to time synchronization errors because of time of flight (TOF) measurements, mmWave radars are more robust in this regard. Therefore, applying this synchronization strategy to FMCW radar systems, which inherently provide Doppler measurement capability, is also well justified from a system-level perspective.

\subsection {Target Positioning: Trilateration}
\label{subsec:static_target_positioning}
In our previous work~\cite{qin_Indoor_2024}, we demonstrated accurate indoor localization using two orthogonally-placed monostatic single-input single-output (SISO) radars. 
Taking FMCW radar as an example, the beat signal resulting from mixing the transmitted signals $s_{\text{tx}}(t)$ and received signals $s_{\text{rx}}(t)$  is derived as
\begin{equation}
    \begin{aligned}
        s_{\text{if}}(t) &=  LPF[s_{\text{rx}}(t) \cdot s_{\text{tx}}(t)]  \\
                     &= \dfrac{A}{2} \cos\left(2\pi k t_d t + \phi_0\right)
    \end{aligned}
\end{equation}
where $LPF$ is a low-pass filter,
     $A$ is the received signal amplitude,
     $t_d = 2r/c$ is the round-trip time delay,
     $r$ is the target range,
     $c$ is the speed of light,
     $k = B/\tau$ is the chirp ramp rate,
     $B$ is the chirp bandwidth,
     $\tau$ is the pulse duration,
     $\phi_0 = -2\pi f_c t_d - \pi k t_d^2$ is a constant phase term, $f_c$ is the carrier frequency.

The range profile is obtained through discrete Fourier transform of the sampled beat signal
\begin{equation}
S[m] = \left| \sum_{n=0}^{N-1} w[n] \cdot s_{\text{if}}[n] \cdot \exp\left(-j \dfrac{2\pi m n}{N}\right) \right|
\end{equation}
where $s_{\text{if}}[n]$ are the discrete samples of $s_{\text{if}}(t)$ at sampling frequency $f_s$,
    $w[n]$ is the window function applied for sidelobe suppression,
    $N$ is the number of samples per chirp,
    $m$ is the range bin index.

Various forms of constant false alarm rate (CFAR) detection can be used for target detection. In cell-averaging CFAR (CA-CFAR), for example, an adaptive threshold $T$ is set according to the following equation 
\begin{equation}
    T = \alpha \cdot Z, Z = \frac{1}{N_{\text{ref}}} \sum_{i=1}^{N_{\text{ref}}} S[m_i]^2
\end{equation}
where 
    $N_{\text{ref}}$ is the number of reference cells,
    $\alpha$ is the threshold scaling factor derived from the desired probability of false alarm ($P_{fa}$).
    The target is declared at the cell under test (CUT) if $S[m_{cut}]^2 > T$.

Assuming that there is one target in the field of view (FoV) of both radars, the CFAR algorithm will output a peak point, and the post-detection measured range between the target and the two radars $r_1$ and $r_2$ are 
\begin{equation}
    r_{1,2} = \dfrac{c m_{1,2}}{2B}
\end{equation}
where $m_{1,2}$ is the range bin index of the two peak points.


For the system shown in Fig.\,\ref{fig:main}(b),  the target's coordinates $(x,y)$ satisfy the equation
\begin{equation} 
    \left\{
        \begin{array}{l}
            (x - a)^2 + y^2 = r_1^2, \\
            x^2 + (y - b)^2 = r_2^2.
        \end{array}
    \right. 
\end{equation}
The orthogonal placement configuration is adopted as opposed to co-lateral or contra-lateral arrangements to mitigate the risk of signal degradation arising from target motion perpendicular to the radar's line of sight~\cite{shen_indoor_2023}. $x$ and $y$ can be solved by trilateration of the distance measurements as
\begin{equation}\label{eq:measurements}
    \left\{
        \begin{array}{l}
            x = \dfrac{a}{2} - \dfrac{a(r_{1}^{2} - r_{2}^{2})}{2(a^2 + b^2)} \pm \dfrac{b}{2}\sqrt{2\dfrac{r_{1}^{2} + r_{2}^{2}}{a^2 + b^2} - \dfrac{(r_{1}^{2} - r_{2}^{2})^{2}}{(a^2 + b^2)^{2}} - 1}, \\
            y = \dfrac{b}{2} + \dfrac{b(r_{1}^{2} - r_{2}^{2})}{2(a^2 + b^2)} \pm \dfrac{a}{2}\sqrt{2\dfrac{r_{1}^{2} + r_{2}^{2}}{a^2 + b^2} - \dfrac{(r_{1}^{2} - r_{2}^{2})^{2}}{(a^2 + b^2)^{2}} - 1}.\\ 
        \end{array}
    \right. 
\end{equation}

(\ref{eq:measurements}) shows the limitations of trilateration, where the $\pm$ operator introduces solution ambiguity, and the GDOP for specific target geometric positions may induce significant amplification.

\subsection {Target Positioning and Tracking: EKF}
When the scene contains moving targets, as often is the case for home and office settings, dynamic tracking of targets becomes valuable. Dynamic target tracking generally takes two steps: 1) state measurement, and 2) estimation~\cite{li_Indoor_2023, Zheng_Catch_2023, guo_2024_uwtracking}. State measurements include target positioning and velocity measurement. Due to the complexity of dynamic scenarios, direct state measurement results are often unreliable or susceptible to noise. The subsequent sections will provide a more detailed discussion of these problems.



In dynamic target positioning, the beat signal is
\begin{equation}
    s_{\text{if}}(t) = \dfrac{A}{2} \cos\left(2\pi (k t_d - f_d) t + \phi_0\right)
\end{equation}
where $f_d = {2 v f_c}/{c}$ is the Doppler frequency shift for a target with radial velocity $v$. The range-Doppler map (RDM) is obtained through a two-dimensional discrete fourier transform of the sampled beat signal across $L$ chirps
\begin{equation}
    \begin{cases}
        S[m, p] = \left| \sum_{l=0}^{L-1} w_2[l] \cdot S[n, l] \cdot \exp\left(-j\dfrac{2\pi p l}{L}\right) \right|, \\
        S[n, l] = \sum_{n=0}^{N-1} w_1[n] \cdot s_{\text{if}}[n, l] \cdot \exp\left(-j\dfrac{2\pi m n}{N}\right),
    \end{cases}
\end{equation}
where
$s_{\text{if}}[n, l]$ are the discrete samples of the beat signal for the $n$-th sample in the $l$-th chirp,
$w_1[n]$ and $w_2[l]$ are the window functions applied for sidelobe suppression in the range and Doppler dimensions,
$L$ is the number of chirps per frame,
$p$ is the Doppler bin index.

The two-dimensional CFAR detection is formulated as:
\begin{equation}
    T[m, p] = \alpha \cdot Z, \quad Z = \dfrac{1}{N_{\text{ref}}} \sum_{i=1}^{N_{\text{ref}}} S[m_i, p_i]^2.
\end{equation}

For each detected target peak at indices $(m_{1,2}, p_{1,2})$ in the RDM, the post-detection radial velocity between the target and the two radars $r_1$ and $r_2$ is 
\begin{equation}
    v_{1,2} = \dfrac{\lambda p_{1,2}}{2LT_{PRI}}
    \label{velocity}
\end{equation}
where $\lambda$ is radar wavelength, $p_{1,2}$ is the Doppler bin index of the two peak points, $T_{PRI}$ is the pulse repetition interval (PRI).


The radial velocity measured by each radar represents the projection of the actual target velocity vector onto the radar's Line-of-Sight (LoS) direction. Geometrically, as illustrated in Fig.~\ref{fig:main}(c), if the actual velocity vector is represented as the diameter of a circle, the radial velocity vectors correspond to the chords starting from the origin and ending on the circle's circumference. This relationship is based on Thales's theorem, which dictates that the projection of a diameter onto any chord passing through its endpoint forms a right-angled triangle. Consequently, the actual velocity vector can be uniquely synthesized by determining the common diameter that accommodates these radial projections.

Based on the above geometric relationships, the actual velocity vector can be described by radial velocity vectors as
\begin{equation}\label{velocity vector}
    \boldsymbol{v} = \dfrac{v_1^2 \boldsymbol{v}_2^{\perp} - v_2^2 \boldsymbol{v}_1^{\perp}}{\boldsymbol{v}_1 \boldsymbol\times \boldsymbol{v}_2}
\end{equation}
where $\boldsymbol{v}_i^{\perp}$ represents the vector $\boldsymbol{v}_i$ rotated 90° clockwise and $\boldsymbol\times$ is the cross product. For various configurations of distributed multi-site radar systems, ~\cite{Zeng_2021_Massive,Ling_2024_Velocity} have proved the actual velocity vector of a target can be derived when its radial velocity vectors are accurately obtained. 

\begin{figure}
    \centering
    \includegraphics[scale=1]{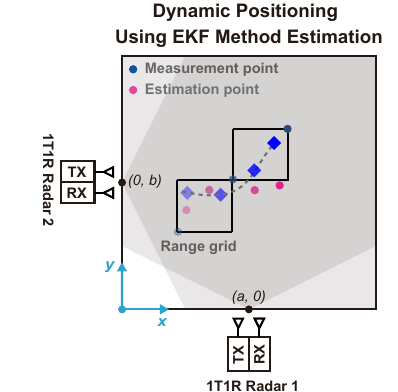}
    \caption{\ Traditional position estimation using EKF method}.
    
    \label{fig:EKF}
\end{figure}

For nonlinear motion prediction, such as random human motion trajectory, the extended Kalman filter (EKF) is often used. EKF optimally combines dynamic model prediction with observation data to mitigate noise effects and provide smoothed trajectory prediction. The fundamental approach involves linearizing the nonlinear system by applying first-order Taylor expansion truncation to the nonlinear function, thereby neglecting higher-order terms. This transformation converts the original nonlinear problem into a linear form suitable for Kalman filtering implementation.
\begin{equation}
    \begin{cases}
        x_k &= f(x_{k-1}) + w_k, \quad w_k \sim \mathcal{N}(0, Q) \\
        z_k &= h(x_k) + v_k, \quad v_k \sim \mathcal{N}(0, R)
    \end{cases}
\end{equation}
where $x_k$ is state vector at time $k$, $x = [p_x, p_y, v_x, v_y]^T$, $p_x$ and $p_y$ are 2D position coordinates, $v_x, v_y$ are velocity components. ${f}(\cdot)$ is state transition matrix, and the system dynamics it represents are typically modeled using the constant velocity (CV) motion model, $w_k$ is process noise with covariance $Q = \mathbb{E}[w_k w_k^T]$ is process noise covariance (model uncertainty). The state transition matrix is typically derived from Newtonian kinematic equations. $z_k$ is observed position measurements, $\bm{z}_k = [z_k^{(1)}, z_k^{(2)}, \ldots, z_k^{(n)}]^\mathrm{T}$, $n$ is the total number of radar, $H$ is observation matrix, $ R = \mathrm{diag}(\sigma_1^2, \sigma_2^2, \ldots, \sigma_n^2) $ is measurement noise covariance, $\sigma_i^2$ denotes the observation noise variance of the $i$-th radar.

Then the EKF linearizes these functions through first-order Taylor expansion.
\begin{equation}
    F_k \approx \left.\dfrac{\partial f}{\partial x}\right|_{\hat{x}_{k-1}}, \quad
    H_k \approx \left.\dfrac{\partial h}{\partial x}\right|_{\hat{x}_k^-}.
\end{equation}

The prediction and update steps become
\begin{equation}
    \begin{cases}
        \hat{x}_k^- &= f(\hat{x}_{k-1}), \\
        P_k^- &= F_k P_{k-1} F_k^T + Q,
    \end{cases}
\end{equation}

\begin{equation}
    \begin{cases}
        K_k &= P_k^- H_k^T (H_k P_k^- H_k^T + R)^{-1}, \\
        \hat{x}_k &= \hat{x}_k^- + K_k (z_k - h(\hat{x}_k^-)), \\
        P_k &= (I - K_k H_k) P_k^-
    \end{cases}
\end{equation}
where $P_k$ is posterior error covariance (estimation confidence), $^-$ denotes a priori estimates (before measurement), $\hat{\cdot}$ indicates estimated quantities, $I$ is $n\times n$ identity matrix, $z_k - H\hat{x}_k^-$ is Innovation (measurement residual), and $K_k$ is optimal blending factor.


Due to state estimation errors and the inherent limitations of the CV motion model ${f}(\cdot)$, the trajectory tracking results exhibit significant position deviations during abrupt trajectory turn, as illustrated in Fig.~\ref{fig:EKF}. Specifically, a sharp turn introduces large unmodeled acceleration components that far exceed the assumed process noise covariance $\boldsymbol{Q}$, leading to severe model mismatch. The filter's prediction becomes unreliable, resulting in estimation lag and potentially covariance collapse.

\section{Proposed Dynamic Positioning and Tracking System using Multi-site Radars}
\subsection {Velocity Synthesis-assisted State Estimation Algorithm}
\label{subsec:Estimation Algorithm}
\begin{algorithm}[b!]
    \caption{VSA Algorithm}\label{alg:VSA}
    \begin{algorithmic}[1]
        \Require 
        \Statex Radar positions $\mathcal{R} = \{(x_1,y_1),...,(x_n,y_n)\}$
        \Statex Initial radial velocity vectors $\mathcal{V}_0 = [v_1^{(0)},...,v_n^{(0)}]$
        \Statex Final radial velocity vectors $\mathcal{V}_t = [v_1^{(t)},...,v_n^{(t)}]$
        \Statex Time interval $t$, grid resolution $M$
        \Statex Threshold parameters $(\epsilon_d, \epsilon_\theta)$
        \Ensure 
        \Statex Target position estimate $\hat{p_0} = (\hat{x}_0, \hat{y_0}), \hat{p_t} = (\hat{x}_t, \hat{y_t})$
        \Statex Target velocity estimate $\hat{{v}} = (\hat{v}_x, \hat{v}_y)$
        
        \State \text{Determine common distance interval}:
        \State $\mathcal{B} \gets \bigcap_{i=1}^n [r_i^{(0)} - \Delta, r_i^{(0)} + \Delta]$ 
        \State Generate 2D grid $\mathcal{G} \gets \text{UniformGrid}(\mathcal{B}, M)$
        
        \For{each grid point $\boldsymbol{p}_k \in \mathcal{G}$}
            \State Velocity pairs set $\Phi_k \gets \emptyset$
            \For{each radar pair $(i,j) \in \{1,...,n\}^2$ where $i \neq j$}
                \State Compute initial velocity $\boldsymbol{v}_{i,j}^{(0)}$ via (\ref{velocity vector})
                \State Compute final velocity $\boldsymbol{v}_{i,j}^{(t)}$ via (\ref{velocity vector})
                \State $\Phi_k \gets \Phi_k \cup \{(\boldsymbol{v}_{i,j}^{(0)}, \boldsymbol{v}_{i,j}^{(t)})\}$
            \EndFor
            
            \State \text{check}:
            \If{$\Phi_k \neq \emptyset$}
                \State $\text{passed} \gets \text{True}$
                \For{each pair $(\boldsymbol{v}^{(0)}, \boldsymbol{v}^{(t)}) \in \Phi_k$}
                    \State $\Delta_d \gets \left| \|\boldsymbol{v}^{(0)}\| - \|\boldsymbol{v}^{(t)}\| \right|$
                    \State $\Delta_\theta \gets \angle(\boldsymbol{v}^{(0)}, \boldsymbol{v}^{(t)})$
                    \If{$\Delta_d > \epsilon_d$ \textbf{or} $\Delta_\theta > \epsilon_\theta$}
                        \State $\text{passed} \gets \text{False}$
                        \State \textbf{break}
                    \EndIf
                \EndFor
                \If{$\text{passed}$}
                    \State $\mathcal{C} \gets \mathcal{C} \cup \{(\boldsymbol{p}_k, \Phi_k)\}$
                \EndIf
            \EndIf
        \EndFor
        
        \State \Return $(\hat{p}, \hat{{v}})$
    \end{algorithmic}
\end{algorithm}



\begin{figure}
    \centering
    \includegraphics[scale=1]{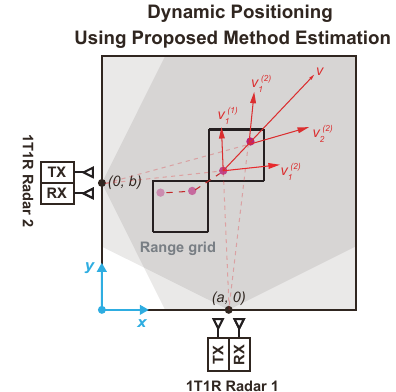}
    \caption{\ Proposed VSA method.
    }
    \label{fig:VSA}
\end{figure}

In this work, we propose the velocity synthesis-assisted (VSA) multi-site radar target state estimation algorithm. The kinematic assumption underlying the proposed VSA method is that, over a short observation time interval $\tau$, the target remains confined to the radar's resolution cell or an adjacent cell. Consequently, it can be regarded as moving with a uniform velocity, as illustrated in Fig.~\ref{fig:VSA}. By leveraging this assumption, the VSA algorithm functions as a robust pre-filtering stage that provides a geometry-validated, instantaneous state measurement $(\hat{\boldsymbol{p}}, \hat{\boldsymbol{v}})$ to the filter, thereby significantly mitigating the reliance on inaccurate predictions and enhancing tracking robustness against high-dynamic maneuvers. 

Algorithm~\ref{alg:VSA} presents the steps of the proposed VSA algorithm. The explanation in detail is as follows.
\paragraph{Spatial Candidate Generation}
The search space is first initialized and discretized. The common distance interval $\mathcal{B}$ is determined as the intersection of all radar range measurements based on (\ref{eq:measurements}), confining the search to the immediate proximity of the target. A uniform 2D grid $\mathcal{G}$ with resolution $M$ is generated over $\mathcal{B}$, treating each point $\boldsymbol{p}_k \in \mathcal{G}$ as a candidate target position.

\paragraph{Multi-Site Velocity Synthesis}
For each candidate position generated in the discretized search space, the algorithm determines the target's velocity vector by exploiting the geometric diversity of the multi-site radar system. Under the assumption that the target maintains a constant velocity during the short observation period, the target state evolution follows a linear kinematic model. Specifically, the positions at the current time $t$ and the previous time $t-1$ are related by
\begin{equation}\label{eq:position_estimation}
    \boldsymbol{p}_t = \boldsymbol{p}_{t-1} + \boldsymbol{v} \Delta t,
\end{equation}
where $\Delta t$ represents the time interval. Based on this instantaneous position, the geometry between the target and the $i$-th radar is characterized by the Line-of-Sight (LoS) vector. This time-varying unit vector is defined as
\begin{equation}\label{eq:los_vector}
    \boldsymbol{u}_{it} = \dfrac{\boldsymbol{p}_t - \boldsymbol{r}_i}{\|\boldsymbol{p}_t - \boldsymbol{r}_i\|},
\end{equation}
where $\boldsymbol{r}_i$ denotes the known coordinates of radar $i$, representing the precise direction of the line-of-sight axis determined by the target and radar positions. Consequently, the radial velocity measured by radar $i$, which corresponds to the projection of the target's full velocity vector onto the LoS direction, is expressed as
\begin{equation}\label{eq:radial_velocity}
    v_{it} = \boldsymbol{v} \cdot \boldsymbol{u}_{it}.
\end{equation}

In the proposed VSA framework, this relationship is pivotal for retrieving the full velocity vector $\boldsymbol{v}$. Since a 2D velocity vector possesses two degrees of freedom, it can be uniquely determined by combining the radial velocity measurements from any two non-collinear radars with overlapping fields of view (FoV). While an orthogonal setup is used in our experiments for optimal coverage, the proposed derivation is generic to any multi-site radar geometry. For every distinct radar pair $(i, j)$, a system of linear equations is constructed based on (\ref{eq:radial_velocity}). By solving this system via matrix inversion, the target velocity is synthesized for both the initial time ($t=0$) and the final time ($t$) using the respective radial velocity measurements $\mathcal{V}_0$ and $\mathcal{V}_t$. This process generates a set $\Phi_k$ of synthesized velocity pairs $(\boldsymbol{v}_{i,j}^{(0)}, \boldsymbol{v}_{i,j}^{(t)})$ that serve as the basis for the subsequent consistency check.

\paragraph{Temporal and Spatial Consistency Check}
This phase enforces the local uniform velocity assumption through a hard constraint check, ensuring that the motion synthesized from all pairs is consistent over the short interval $\Delta t$. This mechanism directly addresses the robustness issue during maneuvers. For every synthesized pair $(\boldsymbol{v}^{(0)}, \boldsymbol{v}^{(t)}) \in \Phi_k$, the change in velocity is quantified by the magnitude difference $\Delta_d$ and the angular difference $\Delta_\theta$. A candidate position $\boldsymbol{p}_k$ is considered valid only if all synthesized velocity pairs satisfy the threshold criteria: $\Delta_d \leq \epsilon_d$ and $\Delta_\theta \leq \epsilon_\theta$. This robust spatial and temporal consistency check effectively filters out candidate positions that are geometrically inconsistent with the target's assumed local inertial motion, thus providing a reliable basis for state estimation even if the true global trajectory involves a sharp turn.

\paragraph{Optimal State Estimation}
The final estimated state $(\hat{\boldsymbol{p}}, \hat{\boldsymbol{v}})$ is derived from the reduced consistent candidate set $\mathcal{C}$. The velocity estimate $\hat{\boldsymbol{v}}$ is typically calculated as a weighted average of the consistent velocity vectors within $\mathcal{C}$, and $\hat{\boldsymbol{p}}$ is selected to minimize the residual error against the current radar measurements. The output $(\hat{\boldsymbol{p}}, \hat{\boldsymbol{v}})$ is then used as the accurate measurement vector $\boldsymbol{z}_k$ input to the EKF update step.


\subsection {Monte Carlo Simulation}
\label{Monte}
The simulation firstly assumes a dual radar system with identical parameters for both radar units. The center frequency of the radars is \SI{60}{GHz}. The bandwidth is \SI{1500}{MHz}, corresponding to a range resolution of \SI{0.10}{\meter}. The frame periodicity is set to \SI{50}{\milli\second}, with $256$ pulses integrated per frame.
We set the SNR to \SI{10}{dB} and the outlier probability to $20\%$ specifically to model indoor environmental conditions. Furthermore, we assumed a synchronization error of \SI{10}{ms} to reflect NTP trigger errors. Radar 1 and Radar 2 are deployed at coordinates $(2.00, 0.00)\,\mathrm{m}$ and $(0.00, 2.00)\,\mathrm{m}$, respectively. A single point target is initialized at position $(1.00, 3.00)\,\mathrm{m}$ with a constant velocity of $(0, -1)\,\mathrm{m/s}$. It is assumed that the target remains within the FoV of both radars throughout the simulation.
The target is situated within the overlapping range bins of both radars, forming a bounded uncertainty region. This region is discretized into a grid of $100$ points for subsequent analysis.

\begin{figure}[t]
    \centering
    \includegraphics{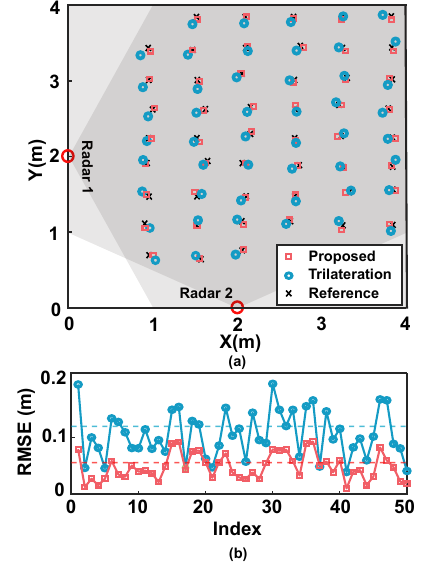}
    \caption{\ Monte-Carlo positioning simulations. (a) Positioning distribution. (b) Positioning errors. } 
    
    \label{fig:Monte-Carlo}
\end{figure}

The positioning performance of the proposed algorithm (Section~\ref{subsec:Estimation Algorithm}) is compared against the conventional trilateration method (Section~\ref{subsec:static_target_positioning}). The robustness of the proposed algorithm is validated through $50$ Monte Carlo simulations, with results shown in Fig.~\ref{fig:Monte-Carlo}. For each simulation, the target position is randomly generated and estimated by both the proposed and conventional trilateration method~(used as the baseline). The proposed algorithm achieved a mean RMSE of $0.06\,\mathrm{m}$, representing a $48\%$ improvement ($0.12\,\mathrm{m}$ $\rightarrow$ $0.06\,\mathrm{m}$) over the baseline method. This comprehensive evaluation demonstrates the algorithm's consistent accuracy and reliability across diverse scenarios.

\begin{figure*}[t]
    \centering
    \includegraphics[scale=0.8]{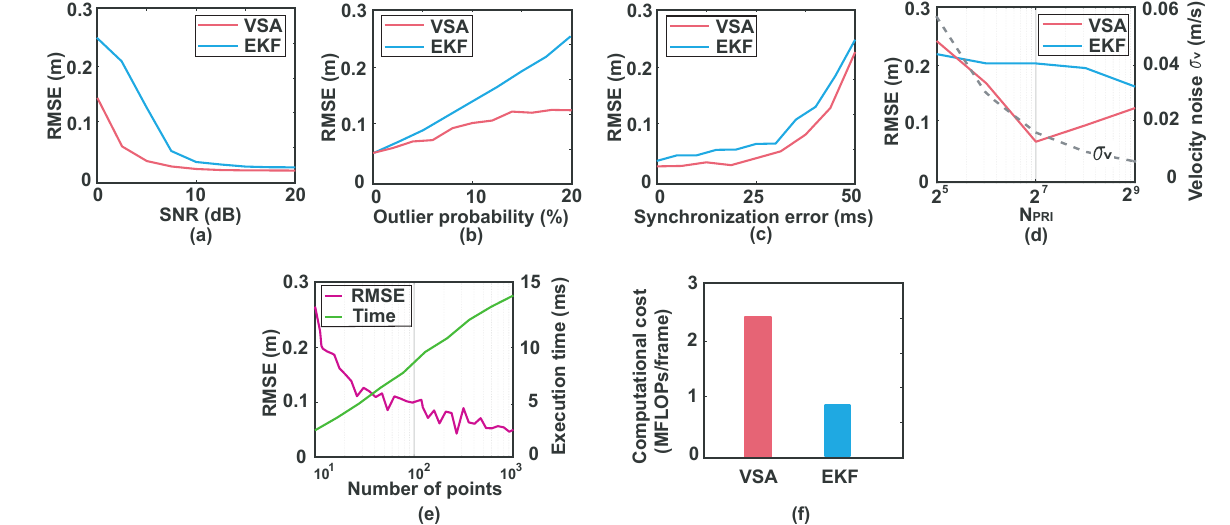}
    \caption{\ Monte-Carlo tracking simulations. (a) SNR. (b) Multipath interference. (c) Inter-radar synchronization latency. (d) The number of PRI. (e) The number of points. (f) Data throughput.}
    
    \label{fig:simulation}
\end{figure*}

To strictly quantify the system's resilience against environmental non-idealities and hardware constraints, extensive Monte Carlo simulations are conducted. Unlike the idealized scenarios before, this analysis explicitly evaluates performance degradation under five critical conditions: low signal-to-Noise ratio (SNR), multipath interference, inter-radar synchronization latency, varying number of points and pulse repetition interval (PRI) configurations. For each scenario, 50 independent Monte Carlo trials are executed to derive statistically significant RMSE metrics.


The tracking accuracy of the proposed VSA method is benchmarked against the conventional EKF over an SNR range of $0$ to $20$~dB. The quantitative comparison, presented in Fig.~\ref{fig:simulation}(a), reflects the theoretical constraint where radar measurement noise variance is governed by the Cram\'{e}r--Rao lower bound (CRLB) and scales inversely with SNR\cite{2025_TMTT_multipath}. Consequently, both methods exhibit performance degradation as the SNR decreases. Nevertheless, the proposed VSA demonstrates significantly enhanced robustness compared to the EKF, particularly in low-SNR regimes. As illustrated, at an SNR of $0$~dB, the EKF yields an RMSE of approximately $0.25$~m, whereas the VSA effectively suppresses the error to $0.14$~m. Although the performance gap narrows as the SNR exceeds $10$~dB, the VSA maintains a consistent accuracy advantage over the conventional approach throughout the tested spectrum.


Indoor environments are characterized by severe multipath effects and antenna non-idealities. Specular reflections and antenna side lobes often generate ghost targets or outliers in the range domain at incorrect angles. To evaluate robustness, an outlier model originating from multipath reflections and side-lobe detections was introduced. Here, the probability $P_{out}$ represents the proportion of measurements corrupted by these anomalies like false echoes, replicas,antenna side lobes and other artifacts relative to the total dataset. Fig.~\ref{fig:simulation}(b) presents the RMSE as a function of outlier probability ($P_{out} \in [0\%, 20\%]$). The conventional EKF lacks an intrinsic mechanism to discriminate between valid maneuvers and large measurement innovations caused by outliers, leading to a linear accumulation of tracking errors. Conversely, the VSA method demonstrates superior resilience, maintaining centimeter-level accuracy even with 20\% outlier contamination. This is achieved through the core spatiotemporal consistency check. False echoes typically produce range measurements that contradict the Doppler signature of the target; the VSA optimization function naturally suppresses these outliers as they fail to satisfy the velocity synthesis condition across the distributed radar network.

In multi-site radar networks using COTS hardware, rigorous clock synchronization is often impractical. For the loose synchronization scheme based on network timestamps, the synchronization error is larger. To verify the robustness of the proposed method against synchronization errors typical of NTP-based networks, a variable time delay $\tau$ ranging from \SI{0}{ms} to \SI{50}{ms} was introduced between the sampling instances of Radar 1 and Radar 2. This range was selected to encompass the worst-case jitter expected in a standard local area network (LAN) environment. The results in Fig.~\ref{fig:simulation}(c) demonstrate that the VSA algorithm maintains robust performance despite these temporal misalignments. Consequently, the proposed NTP-based architecture is validated, demonstrating that strictly wired synchronization is not required to achieve the desired tracking precision.

The number of pulses per frame ($N_{PRI}$) and the pulse repetition interval (PRI) jointly define the coherent processing interval (CPI), such that $T_{CPI} = N_{PRI} \times T_{PRI}$. As shown in Fig. 6(d), increasing $N_{PRI}$ extends the CPI, which refines the Doppler velocity resolution and enhances the VSA estimator's tracking accuracy. This improvement stems from the inverse relationship between Doppler resolution $\Delta v$ and the coherent integration time:
\begin{equation}
    \Delta v = \frac{\lambda}{2 \cdot N_{PRI} \cdot T_{PRI}}
\end{equation}
Consequently, as $N_{PRI}$ increases, the velocity measurement noise $\sigma_v$ (indicated by the grey dashed line in Fig.~\ref{fig:simulation}(d)) decreases significantly. Lower $\sigma_v$ implies a tighter and more accurate velocity constraint for the synthesis algorithm, directly reducing the position estimation uncertainty. 
However, an excessively large $N_{PRI}$ results in a prolonged CPI. Over this extended duration, the constant velocity assumption underlying the Doppler constraint becomes increasingly tenuous due to potential target maneuvering. This kinematic mismatch weakens the efficacy of the velocity constraint, leading to a increase in RMSE. The results indicate that $N_{PRI}=256$ yields the theoretical global minimum in RMSE. This optimum represents a trade-off between enhancing Doppler velocity resolution and preserving the validity of the velocity constraint.

The trade-off between measurement accuracy and computational complexity is investigated by varying the number of grid points. Fig.~\ref{fig:simulation}(e) shows the execution time for each search task (two frames) and RMSE as a function of the number of grid points. All experiments are conducted on an AMD 5800X CPU. As expected, the execution time increases linearly with the number of grid points. The RMSE, however, exhibits diminishing returns. It decreases smoothly but plateaus once the number of grid points exceeds approximately $100$. Therefore, selecting approximately 100 grid points represents a reasonable operating choice. Fig. \ref{fig:simulation}(f) further compares the computational efficiency between the proposed VSA and the conventional EKF at this setting. Both methods are evaluated under the same tracking scenario with identical state dimension and number of targets. It is evident that the VSA incurs a higher computational overhead due to the inherent complexity of the grid search. Comparatively, the VSA prioritizes estimation precision over computational speed, incurring a higher processing overhead than the EKF to achieve robust measurement results. In addition, the extra computational overhead is merely in the order of milliseconds, which barely affects real-time capabilities.





\subsection {Standard Trajectory Simulation}
\begin{figure*}
    \centering
    \includegraphics{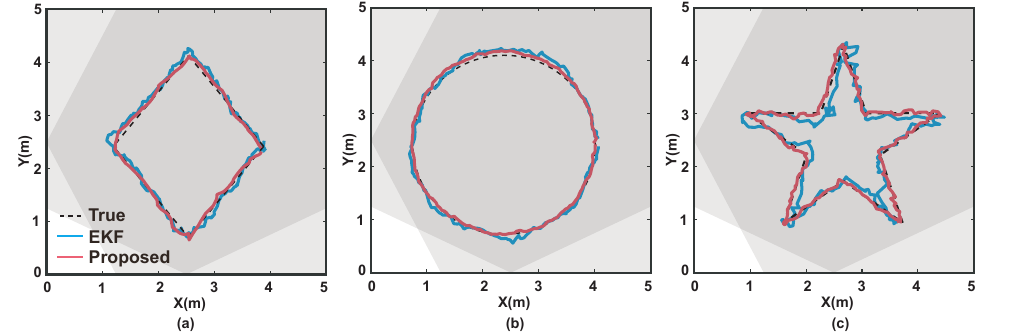}
    \caption{ Simulation results of standard trajectory tracking. (a) Rhombus trajectory. (b) Circular trajectory. (c) Star-shape trajectory.
    }
    \label{fig:standard_trajectory}
\end{figure*}

\begin{figure}
    \centering
    \includegraphics[scale=0.9]{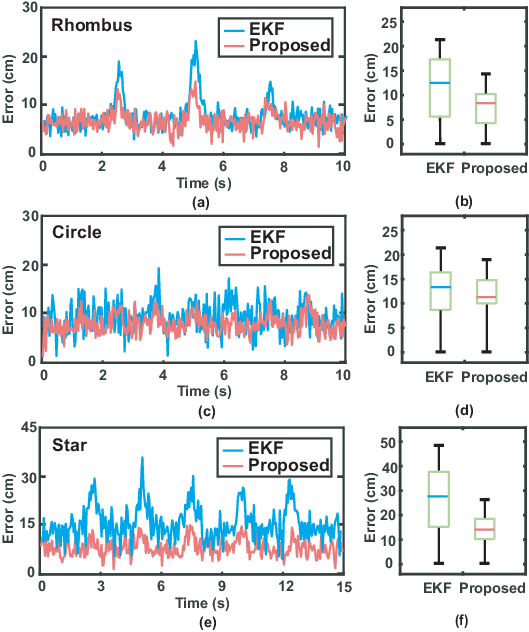}
    \caption{ Comparison of error in EKF and proposed method. (a), (c) and (e) illustrate the variation of error over time. (b), (d), and (f) are error distribution boxplot.
    }
    \label{fig:trajectory_analysis}
\end{figure}
The simulation parameters of radar configuration is the same as before. A point target moves at a velocity $\boldsymbol{v} \in [0.2, 1.5]\,\mathrm{m/s}$, which covers the typical range of human walking speeds from slow strolling to brisk walking. Its speed is reduced when the trajectory changes, ensuring the algorithm's applicability to diverse human motion scenarios. To rigorously evaluate the tracking algorithm under diverse motion patterns and enhance the generalizability of our findings, we designed a set of standard trajectories: rhombus, circular, and star-shaped, as illustrated in Fig.~\ref{fig:standard_trajectory}. These trajectories are chosen to challenge the algorithm and assess its performance across different kinematic scenarios.

Fig.~\ref{fig:standard_trajectory} presents a comparative analysis of tracking performance, contrasting the raw detection points, estimates from a conventional EKF, and the results from the proposed method.
The results indicate that the proposed method achieves satisfactory performance on the relatively regular rhombus and circular trajectories. Notably, it demonstrates a significant performance improvement over the conventional EKF on the complex star-shaped trajectory, where the baseline method struggles with sharp directional changes.

The quantitative performance metrics, detailed in Table~\ref{tab:RMSE}, demonstrate that the proposed method significantly reduces RMSE compared to conventional approaches.
\begin{table}[h]
\centering
\caption{Performance Comparison of Different Trajectory}
\rowcolors{4}{gray!20}{white}
\begin{tabular}{@{}lccc@{}}
\toprule
\midrule
 & \multicolumn{3}{c}{RMSE (m)} \\
\cmidrule(lr){2-4}
 & Rhombus & Circle & Star \\
\midrule
EKF & 0.13 & 0.14 & 0.28\\
Proposed & \textbf{0.09} & \textbf{0.12} & \textbf{0.13}\\
\midrule
\bottomrule
\end{tabular}
\label{tab:RMSE}
\end{table}

Fig.~\ref{fig:trajectory_analysis} presents the position estimation errors for different trajectories. A key observation is that the conventional EKF exhibits a sharp increase in error at corner regions where the target undergoes sudden state transitions, due to model mismatch~\cite{Li_2003_maneuvering,barshalom2004estimation}.
The proposed method effectively mitigates this issue. This performance improvement can be further elucidated from the perspective of Doppler constraints. Unlike the EKF, which relies heavily on the temporal continuity of the kinematic model, the VSA stage exploits the instantaneous radial velocity information embedded in the Doppler shift. The measured Doppler frequency imposes a strict physical constraint on the target's velocity vector, effectively restricting the feasible solution space to a specific manifold. Consequently, even in the absence of accurate prior state predictions during sharp turns, the algorithm effectively rejects candidate locations that violate this Doppler consistency, ensuring robust re-acquisition of the target state measurement.

For the rhombus trajectory, the proposed method yields a more concentrated error distribution than the conventional EKF and effectively suppresses estimation errors at each turning point. For the circular trajectory, the persistent nonlinear motion induces elevated errors in the EKF, however, the inherent stability of this motion pattern results in a diminished relative improvement from our error suppression approach. Most notably, for the complex star-shaped trajectory, frequent and sharp state transitions induce significantly larger estimation errors. The proposed method effectively reduces such errors, achieving superior tracking accuracy with substantially reduced deviations across all challenging motion patterns.


\section{Experimental Validation}
To rigorously validate the effectiveness of the proposed method in practical scenarios, a series of experiments were designed. The study first conducts a comparative assessment of the tracking errors on standard trajectories between a multi-site radar configuration and a MIMO radar system. Subsequently, it examines the performance differences among various multi-site SISO radar methodologies to benchmark and demonstrate the advantages of our approach.

\subsection {Measurement setup}
\label{subsec:setup}
\begin{figure}
    \centering
    \includegraphics[width=3.2in]{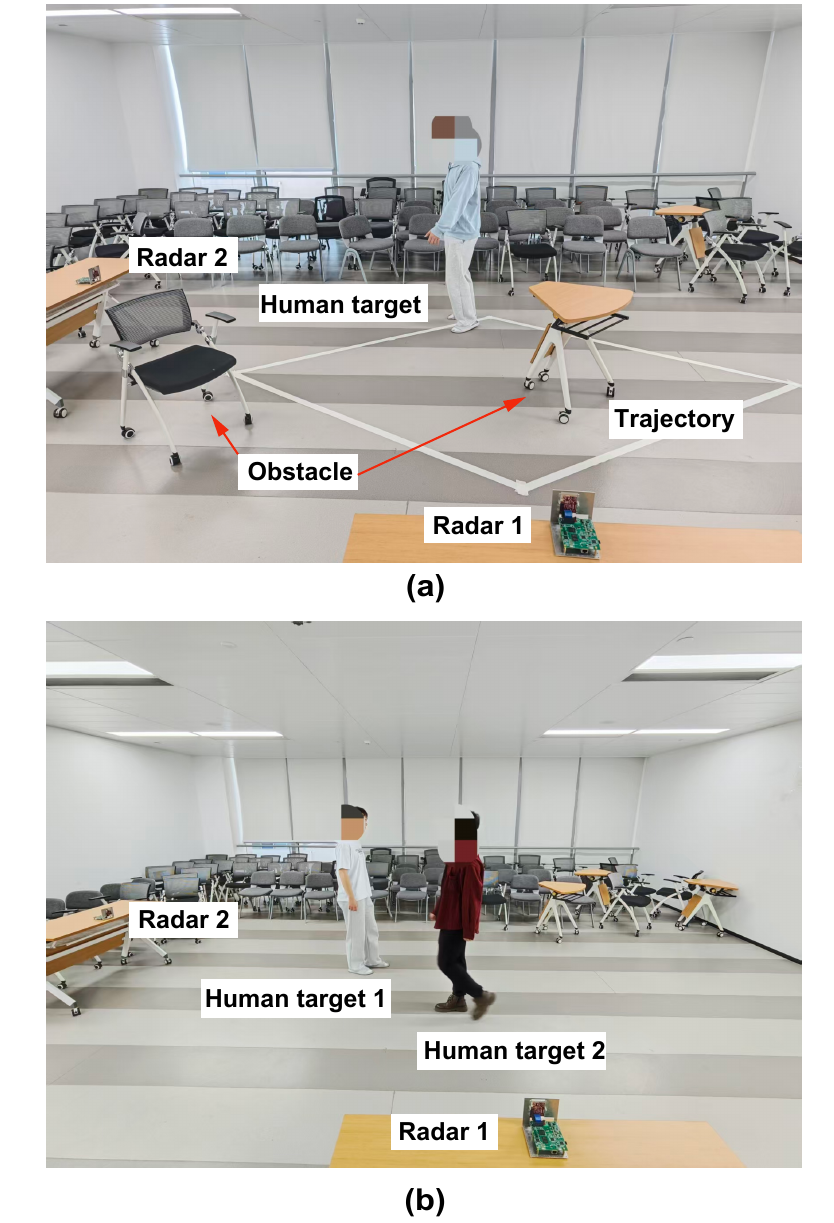}
    \caption{ Experiment setup. (a) A single target with obstacles. (b) Multitarget.  
    }
    \label{fig:setup}
\end{figure}

The experiment was conducted in a $9\,\mathrm{m} \times 9\,\mathrm{m}$ meeting room with the setup shown in Fig.~\ref{fig:setup}. A central measurement area of $5\,\mathrm{m} \times 6\,\mathrm{m}$ was cleared for experiments. The two radars were placed with a relative offset of $2.5\,\mathrm{m}$ horizontally and $3\,\mathrm{m}$ vertically.

The experimental setup used two Texas Instruments AWR6843ISK millimeter wave radar sensors. The waveform is configured with a start frequency of \SI{60}{GHz} and a bandwidth of \SI{1500}{MHz}, corresponding to a range resolution of \SI{0.10}{\meter}. The frame periodicity is set to \SI{50}{\milli\second}, with $256$ pulses integrated per frame, resulting in a maximum unambiguous radial velocity of \SI{3}{m/s}. The detection thresholds for range and Doppler processing were both set to \SI{10}{dB}, with a FoV of $120^\circ$ for angle of arrival estimation.

To minimize mounting errors, custom aluminum alloy brackets were fabricated to constrain the radar tilt angle to within $1^\circ$. The radar's operational mode can be flexibly switched between SISO and MIMO by configuring the number of active transmit and receive antennas. In the MIMO mode used for this experiment, the system achieved an angular resolution of approximately $15^\circ$ in the vertical plane.

Experiments were conducted in both single- and dual-human target scenarios, with the single target case specifically featuring occluded conditions. The analysis is presented in three parts: Section~\ref{subsec:Single}(a) is a comparison between single radar and multi-site radar tracking performances in single target scenario; Section~\ref{subsec:Single}(b) is a comparative evaluation of multi-site radar tracking against different algorithms in single target scenario; Section~\ref{subsec:Multitarget} is an assessment of the proposed algorithm’s efficacy in multitarget experiments.

\begin{figure*}
    \centering
    \includegraphics{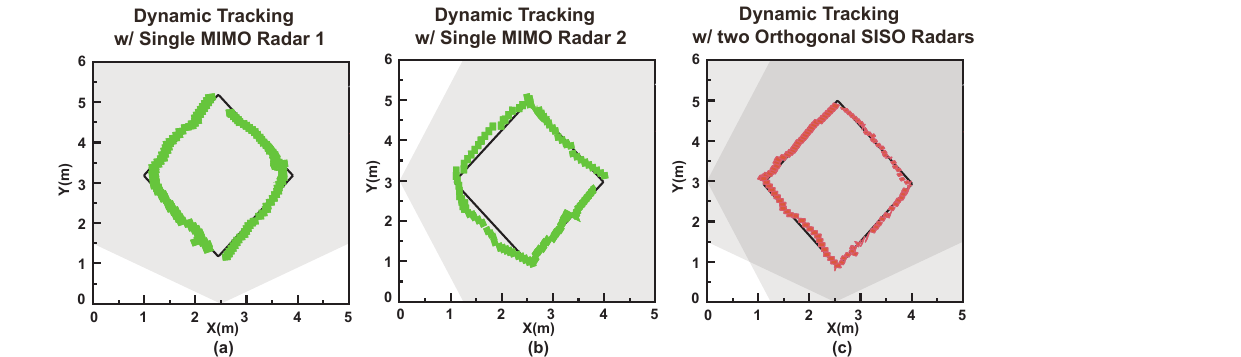}
    \caption{ Experiment results of single radar and multi-site radar trajectory tracking. (a) MIMO radar 1; (b) MIMO radar 2; (c) Multi-site SISO radar; 
    }
    \label{fig:single_radar}
\end{figure*}

\subsection {Single Target Experiments}
\label{subsec:Single}
\paragraph{Assessment of Single Radar and Multi-radar Positioning and Tracking}
This section presents a comparative study between single MIMO radar and the proposed multi-radar approach for indoor positioning and tracking. While a single MIMO radar can theoretically achieve localization due to its inherent angle estimation capability, the human body is typically modeled as an extended target in such systems~\cite{canil_ORACLE_2024,li_Indoor_2023}. MIMO radars often generate a dense set of point cloud, which must be processed using clustering algorithms to extract a representative cluster center.

\begin{table}[ht]
\centering
\caption{Location Comparison of Single Target}
\setlength{\tabcolsep}{3pt}
\rowcolors{4}{gray!20}{white}
\begin{tabular}{@{}lcccc@{}}
\toprule
\midrule
 & \multicolumn{3}{c}{RMSE (cm)} \\
\cmidrule(lr){2-4}
 &  Rhombus  &  circle  &  star \\
\midrule
Radar 1  & 22.4 & 20.2 & 44.2\\
\addlinespace
Radar 2  & 25.8 & 19.8  & 50.1\\
\addlinespace
Point cloud fusion~\cite{li_Indoor_2023}  & 20.4 & 17.6 & 41.0\\
\addlinespace
SAV-PF~\cite{guo_2024_uwtracking} & 16.5 & 15.0 & 22.5\\
\addlinespace
\textbf{This work}  & \textbf{10.5} & \textbf{14.0} & \textbf{16.3}\\
\midrule
\bottomrule
\end{tabular}
\label{tab:single_radar}
\begin{tablenotes}
\small
\item Note: All baseline algorithms were re-implemented and tested in the same experimental scenarios.
\end{tablenotes}
\label{tab:postures}
\end{table}

During the experiments, both radars were initially configured to operate in MIMO mode, with each radar independently tracking the target using an EKF. The systems were then switched to SISO mode, enabling frame alignment through the proposed calibration method in Section~\ref{subsec:setup}, followed by joint trajectory estimation using the algorithm presented in Section~\ref{subsec:Estimation Algorithm}. Participants were instructed to walk along the rhombus trajectory as shown in Fig.~\ref{fig:standard_trajectory}(a) at an approximately constant velocity. 

\begin{figure*}
    \centering
    \includegraphics{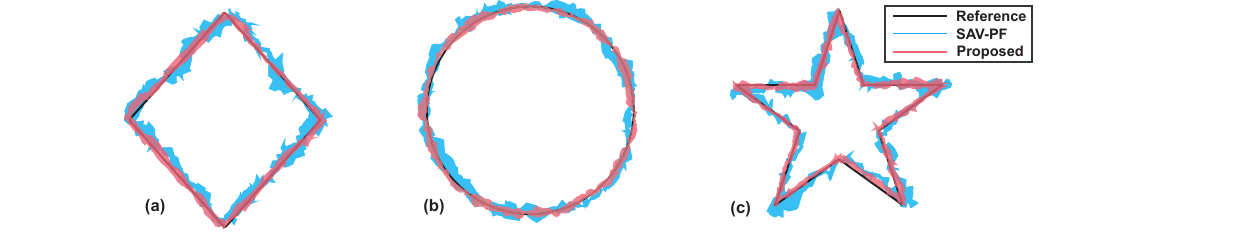}
    \caption{ Experiment results of radar trajectory tracking. (a) Rhombus trajectory.  (b) Circular trajectory.  (c) Star-shaped trajectory.
    }
    \label{fig: final}
\end{figure*}

\begin{figure*}
    \centering
    \includegraphics[scale=0.9]{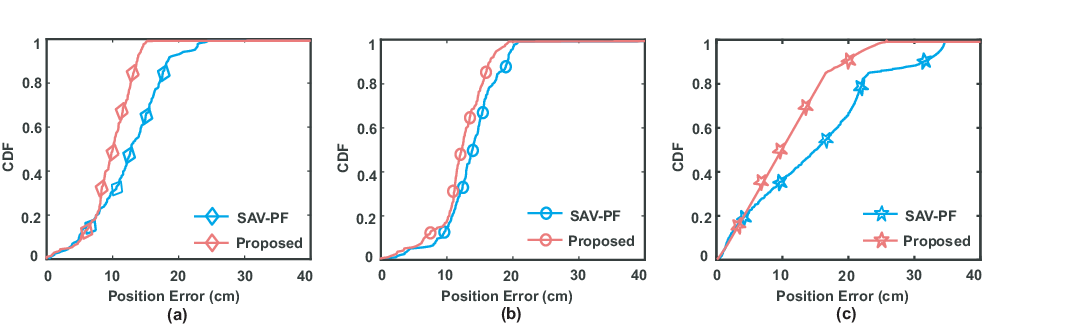}
    \caption{  CDF of estimation error of SAV-PF method and proposed method. (a) Rhombus trajectory.  (b) Circular trajectory.  (c) Star-shaped trajectory.
    }
    \label{fig:CPF}
\end{figure*}

\begin{table*}[ht]
\centering
\caption{COMPARISON OF DIFFERENT WORKS}
\setlength{\tabcolsep}{3pt}
\renewcommand{\arraystretch}{1.5}
\rowcolors{2}{gray!20}{}
\begin{tabular}{lcccccc}
\hline
\hline
\textbf{} & \makecell{\textbf{EKF+tracking fusion}\\~\cite{canil_ORACLE_2024}} & \makecell{\textbf{KF+heatmap fusion}\\~\cite{Zheng_Catch_2023}} & \makecell{\textbf{SAV-PF+heatmap fusion}\\~\cite{guo_2024_uwtracking}} & \makecell{\textbf{KF+point cloud fusion}\\~\cite{li_Indoor_2023}} & \textbf{\textcolor{blue}{This work}} \\
\hline

\begin{tabular}[c]{@{}c@{}}\makecell{Average error \\ (cm)} \end{tabular} & \makecell{\textbf{Single:} 21.0 (Mean) \\ \textbf{Multi:} 25.0(Mean)} & \begin{tabular}[c]{@{}c@{}}\makecell{\textbf{Median error:} \\ \textbf{Single:} 6.0 (Wavy) \\ \textbf{Multi:} 7.0 (Mean)} \end{tabular} & \makecell{\textbf{Single:} \\ 15.2 (Rectangle)\\18.7 (Z-shaped)\\16.4 (Rhombus)\\20.1 (Cross-shaped) \\ \textbf{Multi:} \\ 50 (mean)} & \begin{tabular}[c]{@{}c@{}} \makecell{\textbf{Single:} 8.5 (Zigzag) \\ \textbf{Multi:} 10.0 (Zigzag)}\end{tabular} & \begin{tabular}[c]{@{}c@{}}\makecell{\textcolor{blue}{\textbf{Single:}} \\ \textcolor{blue}{10.5 (Rhombus)}\\\textcolor{blue}{14.0 (Circle)}\\\textcolor{blue}{16.3 (Star)} \\ \textcolor{blue}{\textbf{Multi:}} \\ \textcolor{blue}{13.2 (Side-by-side)} \\  \textcolor{blue}{11.4 (Face-to-face)}} \end{tabular} \\

\begin{tabular}[c]{@{}c@{}}Area (m$\times$m)\end{tabular} & \begin{tabular}[c]{@{}c@{}}\makecell{SISO\\IR-UWB radar}\end{tabular} & \begin{tabular}[c]{@{}c@{}}3$\times$7\end{tabular} & \begin{tabular}[c]{@{}c@{}}5$\times$5\end{tabular} & \begin{tabular}[c]{@{}c@{}}5$\times$7.2\end{tabular} & \begin{tabular}[c]{@{}c@{}}\textcolor{blue}{5$\times$6} \end{tabular} \\
    
\begin{tabular}[c]{@{}c@{}}Radar amount\end{tabular} & \begin{tabular}[c]{@{}c@{}}$2 \sim 4$\end{tabular} & \begin{tabular}[c]{@{}c@{}}2\end{tabular} & \begin{tabular}[c]{@{}c@{}}2\end{tabular} & \begin{tabular}[c]{@{}c@{}}2\end{tabular} & \begin{tabular}[c]{@{}c@{}}\textcolor{blue}{2} \end{tabular} \\

\begin{tabular}[c]{@{}c@{}}Radar model\end{tabular} & \begin{tabular}[c]{@{}c@{}}\makecell{MIMO\\FMCW radar}\end{tabular} & \begin{tabular}[c]{@{}c@{}}\makecell{SISO\\IR-UWB radar}\end{tabular} & \begin{tabular}[c]{@{}c@{}}\makecell{SISO\\IR-UWB radar}\end{tabular} & \begin{tabular}[c]{@{}c@{}}\makecell{MIMO\\FMCW radar}\end{tabular} & \begin{tabular}[c]{@{}c@{}}\makecell{\textcolor{blue}{SISO}\\\textcolor{blue}{FMCW radar}} \end{tabular} \\
        
Range res. (cm) & / & 10 & 10 & 4.4 & \textcolor{blue}{10} \\
\hline
\hline
\end{tabular}
\label{tab:comparison}
\end{table*}

Fig.~\ref{fig:single_radar} illustrates the trajectories reconstructed using EKF from single MIMO radar data. The results are similar to Fig.~\ref{fig:standard_trajectory}-(a) because the human body is an extended target, and the density of the estimated state points are typically lower than that of the point target as modeled in the simulation. It can be observed that a single radar exhibits tracking failures in scenarios involving abrupt trajectory changes. For example, Radar~1 loses track at the top and bottom vertices of the rhombus, while Radar~2 exhibits failures at the right vertex. In contrast, the proposed multi-site SISO radar system produces continuous trajectories with significantly reduced estimation errors, owing to the availability of multi-perspective measurements. These results align with conclusions previously reported in~\cite{li_Indoor_2023}.

To quantitatively evaluate the performance differences, we compare the RMSE obtained by different methods. Table~\ref{tab:single_radar} reports the trajectory estimation RMSE for single MIMO radars under different trajectories, followed by results from direct real-time point cloud fusion. Finally, the proposed method is applied to process the real-time data from both radars without relying on angular measurements. the proposed  method consistently outperforms the Point cloud fusion method approach across all target shapes. Specifically, the RMSE is reduced by $49\%$ (\SI{9.9}{\centi\meter}) for the rhombus trajectory, $20\%$ (\SI{3.6}{\centi\meter}) for the circular trajectory, and $60\%$ (\SI{24.7}{\centi\meter}) for the star-shaped trajectory, highlighting the superior accuracy of the proposed method, particularly in complex trajectories such as the star-shaped trajectory. The results demonstrate that the proposed framework achieves the lowest RMSE, confirming the effectiveness of multi-radar cooperation for robust indoor positioning and tracking. Despite the inherent antenna side lobes present in the commercial radar sensors used, the proposed method maintains high accuracy, further validating the robustness of the VSA algorithm against hardware imperfections.
\paragraph{Assessment of Multi-radar Tracking Between Different Algorithm}
\label{subsec:Multi-radar}

While the previous subsection compared multi-site radar and single MIMO radar tracking performance, this section focuses on analyzing algorithmic differences under identical radar setups. Prior work in~\cite{guo_2024_uwtracking} employs a radar architecture similar to ours and proposes an enhanced particle filtering method that incorporates velocity information and scanning angle to improve tracking accuracy in non-Gaussian noise environments. Specifically, the scanning angle and velocity optimized particle filter (SAV-PF) method derives the target’s velocity components by differencing its historical trajectory and generates candidate particles through dynamic perturbations, including angular and velocity perturbations. These particles are subsequently refined using particle filtering, where those best satisfying motion continuity constraints are selected for trajectory estimation.


In target tracking scenarios, errors may originate from multiple sources. To enable a fair comparison of algorithm performance, we use error bands as a baseline reference. The error bands are obtained from five repeated measurements, which effectively mitigate the randomness of a single trial and provide a more intuitive and compelling visual comparison. Fig.~\ref{fig: final} illustrates the tracking results for the proposed standard trajectories with corresponding error bands, and the corresponding cumulative distribution function (CDF) comparison under different trajectory patterns is presented in Fig.~\ref{fig:CPF}. It can be observed that the proposed method consistently outperforms the SAV-PF algorithm in error suppression. This advantage arises because the proposed algorithm directly extracts accurate positional and velocity information of the target and employs it as the predictive output of the current state. In contrast, SAV-PF relies primarily on historical trajectory information for prediction, which leads to delayed response when facing abrupt trajectory changes and results in error accumulation. The RMSE for proposed method and SAV-PF method are in the Table~\ref{tab:single_radar}. The median tracking errors for proposed method and SAV-PF method are \SI{10.5}{\centi\meter} and \SI{16.5}{\centi\meter} in rhombus trajectory, \SI{14.0}{\centi\meter} and \SI{15.0}{\centi\meter} in circular trajectory, \SI{16.3}{\centi\meter} and \SI{22.5}{\centi\meter} in star-shaped trajectory. These results clearly demonstrate that the proposed method exhibits superior robustness across different trajectory patterns.


\subsection {Multitarget Experiments}
\label{subsec:Multitarget}
We test trajectory tracking of two targets moving in parallel at distances of \SI{1.5}{m} as shown in Fig.~\ref{fig:standard_trajectory}(b). In a range-only trilateration system lacking phase coherence, the true pairing of range measurements from different radars is unknown, leading to the generation of false (ghost) targets alongside real ones. However, in scenarios characterized by non-intersecting trajectories, velocity information can be leveraged to associate true target tracks. While multi-target separation remains a complex challenge, this work presents some preliminary measurement results. Specifically, in scenarios involving two subjects walking side-by-side and face-to-face directions, the mean positioning errors derived from multiple measurements were \SI{13.2}{cm} and \SI{11.4}{cm}, respectively.

\subsection {Comparison with existing works}
A performance comparison between the proposed system and existing references in single target situation is summarized in Table~\ref{tab:comparison}. In~\cite{canil_ORACLE_2024}, the multiple object tracking performance precision (MOTP) was evaluated for various trajectories, including in-line, parallel, circular, and free movements. The study in~\cite{Zheng_Catch_2023} analyzed median tracking distance error under wavy trajectories using parallel radar architectures. Building on this,~\cite{guo_2024_uwtracking} extended the evaluation to multiple trajectory types and proposed the SAV-PF algorithm. In addition,~\cite{li_Indoor_2023} analyzed average error under linear zigzag trajectories using orthogonal radar architectures. 

Although the lack of unified standards for evaluating identical error metrics under consistent test trajectories makes direct comparison with existing works challenging, our method demonstrates superior performance. In single-target experiments, it successfully tracks the star-shaped trajectory, which is significantly more complex than the paths typically utilized in current literature. Notably, for the rhombus trajectory in the single-target scenario, our approach reduces the root-mean-square error (RMSE) by $36\%$ (\SI{5.9}{\centi\meter}) compared with the SAV-PF method in~\cite{guo_2024_uwtracking}. Furthermore, the proposed system maintains consistently low error levels in multi-target scenarios.

\section{Conclusion}

In this work, we proposed a multi-site SISO radar architecture that eliminates the need for expensive multi-channel transceivers and strict hardware synchronization. By utilizing a practical NTP trigger and spatially distributed single-channel nodes, the system significantly simplifies hardware complexity while capturing diverse spatial information. Central to this approach is the novel velocity synthesis-assisted (VSA) algorithm, which exploits Doppler velocity information to resolve positioning ambiguities. Crucially, the VSA algorithm exhibits superior robustness against time alignment errors and multipath environment, enabling high-precision sensing in realistic environments without intricate synchronization facilities. Experimental results demonstrate that the system achieves centimeter-level tracking accuracy, outperforming state-of-the-art solutions in complex trajectory scenarios. Consequently, this work presents a promising, low-cost solution for long-term human monitoring, with future research directed towards multi-target tracking capabilities.

\section*{Acknowledgment}
This work is supported by the Shenzhen Science and Technology Program under Grant (JCYJ20230807091814030, JCYJ20220818100408018 and 20231115204236001), the National Natural Science Foundation of China under Grant (62471211 and 32371992), and in part by the Guangdong Basic and Applied Basic Research Foundation under Grant (2025A1515011109 and 2024A1515011902).

\ifCLASSOPTIONcaptionsoff
  \newpage
\fi



~
\bibliographystyle{IEEEtran}
\bibliography{lumped}

\vfill


\end{document}